\documentclass[10pt,conference,a4paper,hidelinks]{IEEEtran_EDM}
\IEEEoverridecommandlockouts
\usepackage{cite}
\usepackage{amsmath,amssymb,amsfonts}
\usepackage{algorithmic}
\usepackage{listings}
\usepackage{graphicx}
\usepackage{textcomp}
\usepackage{xcolor}
\usepackage{hyperref}
\def\BibTeX{{\rm B\kern-.05em{\sc i\kern-.025em b}\kern-.08em\kern-.1667em\lower.7ex\hbox{E}\kern-.125emX}}

\hypersetup{draft}

\def\confheader{}

\makeatletter
\newcommand{\linebreakand}{%
  \end{@IEEEauthorhalign}
  \hfill\mbox{}\par
  \mbox{}\hfill\begin{@IEEEauthorhalign}
}
\makeatother

\usepackage{flushend} 

\lstdefinelanguage{reflex}
{
    morekeywords={set,state,stop,next,start,process,error,timeout,reset,timer},
	morecomment=[l]{//},
	basicstyle=\ttfamily,
}

\definecolor{codeblue}{rgb}{0.13,0.67,0.8}
\lstdefinestyle{IsabelleStyle}{
    language=Isabelle,
    keywordstyle=[1]\color{codeblue},
    keywordstyle=[2]\color{black},
    keywordstyle=[3]\bfseries
}
\lstdefinelanguage{Isabelle}{
    keywords=[1]{datatype,type_synonym,and,definition,if,then,else,case,inductive},
    keywords=[2]{nat,int,real,bool,byte,word,dword,lword,string,wstring,time},
    keywords=[3]{=,=>,|},
   morecomment=[s]{"}{"}
}
\begin{document}

\markboth{\confheader}{}
\title{
\mbox {Partial Automation of Verification} Condition Proving for Reflex Programs (Draft)

}
\author{
\IEEEauthorblockN{
Artyom Ishchenko
}
\IEEEauthorblockA{
\textit{Novosibirsk State University,}\\
\textit{Department of Computer Science Systems}\\
\textit{Institute of Automation and Electrometry,}\\
\textit{Laboratory of Cyberphysical Systems}\\
Novosibirsk, Russia \\
0009-0005-6176-5917 
}
\and
\IEEEauthorblockN{Igor Anureev
}
\IEEEauthorblockA{
\textit{Institute of Automation and Electrometry,}\\
\textit{Laboratory of Cyberphysical Systems}\\
Novosibirsk, Russia \\
0000-0001-9574-128X 
}
}
\maketitle
\blfootnote{This work was supported by the Russian Ministry of Education and Science, project 125022803031-1}
\begin{abstract}
Process-Oriented Programming is a software development approach that emphasizes the management of control systems through abstractions of processes and their states, enabling these systems to be described in terms of real physical processes. This native description of control is particularly important for industrial systems consisting of hundreds or thousands of processes. For such systems, safety is critical. To ensure the reliability and safety of these systems, formal verification methods must be applied. One such method is deductive verification, which involves formalizing programs and their requirements as logical formulas, known as verification conditions. Proving these conditions confirms that the program meets its requirements. The automatic generation of verification conditions is performed by a specialized software tool called a verification condition generator. We previously proposed a verification condition generator for the Reflex language. However, it generates too many verification conditions, making their manual proof impossible. This paper proposes modifications to the verification condition generator aimed at automating the proof of some of these conditions. These modifications include introducing an annotation language to describe requirements in a structured form, generating invariants based on the program structure, and using SMT solvers for the preliminary attempt to solve the verification conditions.
\end{abstract}


\section{Introduction}

The process-oriented paradigm \cite{zyubin2007hyper} offers a promising approach to programming industrial control systems. In this paradigm, a program consists of a set of interacting processes, each defined by a sequence of states that specify its execution logic.

The Reflex language \cite{zyubin2018reflex} is a process-oriented dialect of the C programming language, accompanied by a dedicated translator \cite{Bastrykina2020}. Reflex combines the advantages of the process-oriented programming model with the familiar syntax and programming style of C, making it convenient for engineers developing industrial control applications.

Since Reflex programs are used in industrial environments, they must satisfy strict safety and correctness requirements. To ensure these properties, we employ deductive verification \cite{hahnle2019deductive}. In this approach, a program together with its specification is translated into a logical formula expressed in a formal logic. Using a logical inference system -- commonly referred to as the axiomatic semantics of the programming language -- this formula is reduced to a set of logical formulas known as verification conditions (VCs). If all generated verification conditions are proved to be true, the program is considered correct with respect to its specification.

Axiomatic semantics have been developed for many programming languages, including Java \cite{von2001hoare}, abstract imperative languages \cite{schirmer2005verification}, and C \cite{10.1145/2578855.2535878}. In such systems, the formalization of requirements typically results in preconditions and postconditions describing the behavior of program fragments. Various strategies exist for reasoning about these conditions. In this work, we employ the strongest postcondition strategy, which processes the program sequentially from beginning to end and derives postconditions by applying inference rules that transform the current program state.

A practical challenge in deductive verification is the large number of verification conditions generated for realistic programs. In our previous work \cite{ishchenko2025verification}, we developed a verification condition generator (VCG) for Reflex programs. However, when applied to large programs, the tool produced thousands of verification conditions, making manual proof of all of them impractical.

In this paper, we present three extensions to the previously developed VCG aimed at increasing the level of automation in proving verification conditions:
\begin{enumerate}
\item an annotation language for specifying program properties,
\item a structure-based invariant generation mechanism,
\item integration with SMT solvers for automated reasoning.
\end{enumerate}

\section{Preliminaries}
This section describes the Reflex language and the modifications made to it. It then provides an overview of SMT solvers.

\subsection{The Reflex Language}

Reflex is a process-oriented dialect of the C programming language designed for programming industrial logic controllers. The language largely preserves the syntax of C statements and expressions while introducing several adaptations required for control applications. In particular, variable declarations are restricted to specific locations, data types have strictly defined sizes, and additional primitive types such as \textit{time} and \textit{bool} are supported. Reflex also introduces constructs specific to the process-oriented programming paradigm.

A Reflex program consists of a declaration of an activation interval, declarations of variables and constants, and a sequence of process definitions. The activation interval specifies the fixed time duration of a single program iteration within the control loop.

Each process is defined by its local variables and a set of \textit{active} states. In addition, every process has two predefined \textit{inactive} states, \textit{stop} and \textit{error}, and a clock that records the time the process has spent in its current state. Initially, all processes except the first one start in the \textit{stop} state. Processes that are currently in inactive states are considered inactive.

A state represents a named sequence of C-like statements. These sequences may contain standard C statements except loop constructs and the \texttt{goto} statement. In addition, Reflex provides process-oriented statements for changing the state of a process (\textbf{set state}), resetting the process timer (\textbf{reset timer}), and controlling process execution (\textbf{start}, \textbf{stop}, and \textbf{error}). A state may also include a \textit{timeout} construct that checks whether the time spent in the current state exceeds a specified value and executes a corresponding handler if the timeout occurs.

Program execution proceeds in iterations. During each iteration, all processes of the program are executed in textual order according to their current states.

Reflex also provides constructs for expressing lightweight intermediate execution points within a state. The construct \texttt{wait(C)} suspends the execution of the subsequent statements until the Boolean condition $C$ becomes true. For example, if the state body has the form $S_1; wait(C); S_2$, where $S_1$ and $S_2$ are sequences of statements, then $S_1$ is executed first. If the condition $C$ holds, execution proceeds with $S_2$; otherwise, the execution of $S_2$ is postponed until a later program iteration in which $C$ becomes true.

The construct \texttt{slice} separates the execution of a state body across program iterations. If a state body has the form $S_1; slice; S_2$, the statements in $S_1$ are executed during the current iteration, while the execution of $S_2$ continues in the following iteration.

The language also supports structured data types such as arrays and structures, which are declared using a syntax similar to that of C.

The Reflex language implementation is based on the ANTLR4 toolset \cite{10.5555/2501720}. The complete grammar of the language is publicly available online \cite{Ishchenko2024}.

\subsection{Annotation Languages}

Annotation languages are commonly used in deductive program verification to express formal specifications of program behavior directly within the source code. Annotations typically describe properties such as preconditions, postconditions, invariants, and assertions that capture the intended behavior of program fragments. These specifications are not executed as part of the program itself; instead, they serve as logical constraints that guide the verification process. During verification, annotations are translated into logical formulas and combined with the formal semantics of the programming language to generate verification conditions \cite{hahnle2019deductive}.

Many modern verification frameworks rely on dedicated annotation languages or annotation extensions of existing programming languages. For example, the Java Modeling Language (JML) provides specification constructs for Java programs \cite{leavens2006jml}, while ACSL (ANSI/ISO C Specification Language) is used to annotate C programs with formal specifications \cite{baudin2021acsl}. Such languages allow developers to express correctness properties in a structured and machine-readable form, enabling automated reasoning tools to verify that the program satisfies its specification.

The Reflex language already includes an annotation language introduced in \cite{anureev2019two}. However, it lacks sufficient expressiveness and forces engineers to engage with low-level details of the translation logic. This increases the likelihood of errors in requirement specification and limits the kinds of behavior it can describe. Therefore, the annotation language requires redesign.

\subsection{SMT solvers}

Satisfiability Modulo Theories (SMT) solving is a fundamental technique for automated reasoning about logical formulas that arise in program verification, formal methods, and symbolic analysis. SMT extends propositional satisfiability (SAT) by allowing formulas to contain symbols interpreted in background theories such as linear arithmetic, arrays, bit-vectors, algebraic datatypes, and uninterpreted functions \cite{barrett2018satisfiability}. These theories enable a natural encoding of program constructs such as variables, memory structures, and control-flow conditions.

Formally, given a first-order formula $\varphi$ and a background theory $T$, an SMT solver determines whether $\varphi$ is \emph{$T$-satisfiable}, i.e., whether there exists a model that satisfies $\varphi$ while respecting the semantics of the theory $T$. If such a model exists, the solver returns \texttt{sat} together with a satisfying assignment. If no such model exists, the solver reports \texttt{unsat}. When the solver cannot determine the satisfiability of the formula due to incomplete reasoning procedures or resource limits, the result may be \texttt{unknown}.

Modern SMT solvers rely on a combination of Boolean reasoning and theory-specific reasoning procedures. The dominant architecture is based on the \emph{DPLL(T)} framework \cite{nieuwenhuis2006solving}, which integrates a SAT solver with dedicated theory solvers. In this architecture, the SAT solver handles the Boolean abstraction of the formula, while the theory solvers check the consistency of assignments with respect to the background theories and generate additional constraints when inconsistencies are detected.

SMT solvers support a wide range of background theories that are particularly useful in program verification. For example, theories of arrays and algebraic datatypes allow modeling of memory structures and recursive data structures, while uninterpreted functions provide an abstraction mechanism for reasoning about program functions without fully specifying their semantics. Many SMT solvers also support reasoning with quantified formulas, although quantifier handling typically relies on heuristic instantiation techniques and is not complete in general \cite{reynolds2013quantifiers}.

To ensure interoperability between verification tools and SMT solvers, the SMT-LIB initiative provides a standard input language and a large collection of benchmarks \cite{barrett2010smtlib}. SMT-LIB defines a uniform syntax for declaring sorts, functions, and logical constraints, allowing verification conditions generated by analysis tools to be processed by different solvers without modification.

In deductive program verification, SMT solvers are typically used to check \emph{verification conditions} generated from annotated programs and program invariants. Program semantics and correctness properties are translated into logical formulas, and the solver is used to determine whether a violation of the property is possible. If the conjunction of the program semantics and the negation of the desired property is unsatisfiable, the property is considered proven. Due to their efficiency and expressive power, SMT solvers have become a central component of modern verification frameworks. Widely used solvers include Z3 \cite{de2008z3}, CVC4 \cite{barrett2011cvc4} and CVC5 \cite{barbosa2022cvc5}, which support a broad range of theories and advanced solving techniques.

\section{Annotation Language}
Annotations are written as C-style comments and are distinguished by enclosing the content in square brackets. Each annotation consists of a type \textit{T}, an optional logical language specifier \textit{L}, and the content \textit{C}:

\begin{verbatim}
//[T(L): C]
/* [T: C]
[T: C]
*/
\end{verbatim}
The optional language specifier defines the syntax used for the annotation content. If omitted, the default language described in \ref{ann-cont} is assumed. This mechanism allows annotations to be expressed in different formal languages, such as Isabelle/HOL.

Annotations are attached to the first statement or declaration that follows them and are interpreted in the context of that program element.

\subsection{Annotation types}

The following types of annotations are supported:

\begin{itemize}
\item \lstinline{[assume:C]} -- a logical condition that is assumed to hold before the execution of the associated statement. For input variables, it acts as a specification and is taken as given without verification.
\item \lstinline{[assert:C]} -- a logical condition that must hold after the execution of the associated statement. Its content \textit{C} is a formula.
\item \lstinline{[invariant:C]} -- a logical condition that must hold both before and after execution. When attached to a loop, it provides a loop invariant. When attached to a program, process, or state declaration, it contributes to the global invariant of the system. Its content \textit{C} is a formula.
\item \lstinline{[define:C]} -- introduces a new variable, constant, or function. Its content \textit{C} follows the syntax of a C-style definition.
\end{itemize}

Program, process, and state declarations contain sequences of statements that execute within the activation loop. Therefore, their behavior is best described using invariants, and the use of \textbf{assume/assert} annotations in these contexts should be avoided.

\subsection{Annotation content}\label{ann-cont}

The content of an annotation depends on its type, but most annotations involve formulas. A formula is an expression of Boolean type. Constants, variables, and function calls are all valid expressions. The operators $\{+, -, *, /, \%, >>, <<, |, \&, \wedge , ==, !=, <, >, <=, >=, !\}$ are used to construct expressions from other expressions. The operators $\{++, --, +=, -=, *=, /=, \%=\}$ may be used to construct expressions involving annotation-defined variables, but only at the top level (i.e., expressions like $x = y++$ are not permitted). These two groups of operators follow standard C semantics. The logical operators ${\&\&, \mid\mid}$ are also defined, with their usual meanings.

Let $\phi$ be an arbitrary formula, $\rho$ a process name, $\xi$ a process state name, and $\iota$ an expression. The following operators are defined:

\begin{itemize}
\item $\phi_1==>\phi_2$ -- logical implication.
\item $\phi_1<==>\phi_2$ -- logical equivalence.
\item prev($\phi$) -- a predicate stating that $\phi$ was true in the previous iteration of the algorithm.
\item before($\phi$) -- a predicate stating that $\phi$ was true in some previous iteration of the algorithm.
\item between($\phi_1$,$\phi_2$,$\phi_{mid}$) -- a predicate stating that $\phi_{mid}$ holds during the interval between iterations where $\phi_1$ and $\phi_2$ are true.
\item next($\phi$) -- a predicate stating that $\phi$ will be true in the next iteration of the algorithm.
\item eventually($\phi$) -- a predicate stating that $\phi$ will become true in some future iteration.
\item in($\rho$,$\xi$) -- a predicate stating that process $\rho$ is currently in state $\xi$.
\item time() -- a function that returns the amount of time elapsed in the current state of the process.
\item \textbackslash FA($v_1, \dots, v_n$.$\phi$($v_1 \dots v_n$)) -- the universal quantifier, with bound variables $v_1 \dots v_n$ and formula $\phi$.
\item \textbackslash EX($v_1, \dots, v_n$.$\phi$($v_1 \dots v_n$)) -- the existential quantifier, with bound variables $v_1 \dots v_n$ and formula $\phi$.
\end{itemize}

Variables appearing in annotations are interpreted after the associated statement (or declaration) has initialized its variables and after name shadowing has been resolved.

The interpretation of the temporal operators (\textit{prev}, \textit{before}, \textit{between}, \textit{next}, \textit{eventually}) depends on the statement to which the containing expression is attached. For instance, in a program-level invariant, prev refers to the previous program iteration. In a process state invariant, it refers to the previous program iteration, but only if the process was in the same state. For a loop invariant, it refers to the previous loop iteration.

An expression may also specify a scope using the syntax \textit{.scope($\Phi$)}, where $\Phi \in {prev(), before(\phi), next(), eventually(\phi)}$. For example, in $(x+y).scope(before(x<y))$, the expression $x+y$ is evaluated at the first point in the past when $x<y$ held.

The annotation content language described in this section is the default. If no language is explicitly specified, this language is used. Otherwise, the content must conform to the syntax of the specified language. For example, an annotation using Isabelle/HOL would be written as:
\begin{verbatim}
//[invariant(Isabelle):
    \<forall>s. toEnvP s]
\end{verbatim}
Formulas provided in an external language are not transformed during VC generation, except for variable name mangling to avoid conflicts. At the start of the verification condition generation process, these annotations are translated into the system's internal representation of formulas.

\section{Extra Invariants}\label{extra}
Invariants are a crucial component of verification, as they allow reasoning about program behavior that extends beyond the current iteration. In addition to the invariants provided by the engineer, further invariants can be derived from the program structure and the sequences of statements it contains. These automatically generated invariants are formulated as logical formulas, as defined in \cite{ishchenko2025verification}.

Let $\rho$ be an arbitrary process name, $\xi$ a process state name, and $\nu$ a variable name. The following kinds of structural invariants are supported:
\begin{itemize}
\item \textbf{Process states} -- characterizes the set of states each process can occupy. These invariants have the form $\forall s. toEnvP\ s \longrightarrow getPstate\ s\ \rho \in {\xi_1 \dots \xi_n}$.
\item \textbf{Defined stabilized process state variables} -- specifies the values that variables hold in stabilized process states (states that have been active for more than one consecutive iteration). They are formulated as $\forall s. toEnvP\ s \land getPstate\ s\ \rho = \xi \land getPstate\ (predEnv\ s)\ \rho = \xi \longrightarrow getVarValue\ s\ \nu_1 = \textbf{val}_1 \land \dots \land getVarValue\ s\ \nu_n = \textbf{val}_n$.
\item \textbf{Defined non-stabilized process state variables} -- specifies the values that variables hold upon entering a particular process state. These invariants have the form $\forall s. toEnvP\ s \land getPstate\ s\ \rho = \xi \longrightarrow getVarValue\ s\ \nu_1 = \textbf{val}_1 \land \dots \land getVarValue\ s\ \nu_n = \textbf{val}_n$.
\item \textbf{Process state transition condition} -- describes the condition that must be satisfied to transition into a given process state. These invariants are expressed as $\forall s. toEnvP\ s \land getPstate\ s\ \rho = \xi \longrightarrow \textbf{condition}_1 \lor \dots \lor \textbf{condition}_n$. The conditions are defined using an auxiliary function $prevProcState$, which takes a program state and a process name and returns the most recent program state in which the process's state differed from its current one. This is represented as an intermediate variable using Isabelle/HOL's \textit{let ... in ...} construct. Two types of transition conditions are distinguished:    
\begin{itemize}
\item \textbf{Initial state} has the form $prevProcState\ s\ pname = emptyState$, indicating that the process has been in this state since the start of the program.
\item \textbf{Conditional transition} is a more general form expressed as $let\ s_2 = (prevProcState\ s\ pname)\ in\ P(s_2)$, where $P$ is some predicate. This includes the following subtype:
\begin{itemize}
\item \textbf{Timed transition} takes the form $let\ s_2 = (prevProcState\ s\ pname)\ in\ \newline(ltime\ s_2\  pname\ \ge time) \land P(s_2)$, where $ltime$ returns the current time of process $pname$ and $P$ represents any additional condition.
\end{itemize}
\end{itemize}
\end{itemize}

These invariants are relatively simple and structurally derived; they require only the tracing of program paths, a process similar to that already performed during verification condition generation.

\textbf{Process state} invariants are generated by analyzing the set of state names defined for each process.

\textbf{Defined stabilized process state variables} invariants are derived by examining the sequence of statements within a state. If a variable is assigned the same value at the end of every iteration of that state, it is included in the invariant.

\textbf{Defined non-stabilized process state variables} invariants are constructed in two steps:
\begin{enumerate}
\item Identify all variables that are defined before the state is first entered.
\item Verify that these variables are not modified while the state is active.
\end{enumerate}

Both stabilized and non-stabilized variable invariants also require that the relevant variables are not changed by any processes that execute after the one being analyzed.

\textbf{Process state transition condition} invariants are the most complex, as they require tracing all conditions and variable modifications along the path leading to a \textit{set state} statement.

The \textbf{initial state} condition is added if the process is active from the first iteration of the program and the state being analyzed is its initial state.

A \textbf{conditional transition} condition corresponds to each \textit{set state} statement. Analogous to the verification condition generation process, a list of formulas is accumulated as the process state is traversed. When a \textit{set state} statement is encountered, the collected formulas are filtered to retain only those relevant to the conditional expressions that guard the transition (e.g., conditions from \texttt{if-else} or \texttt{switch} statements). These relevant formulas are then conjoined to form the \textbf{condition}.

A \textbf{timed transition} condition is added when a transition occurs within a \textit{timeout} statement. It follows the same rules as a general \textbf{conditional transition}.

\section{SMT preparation}
To use SMT solvers for checking verification conditions, two steps must be performed:
\begin{enumerate}
\item Translate each verification condition into the input language of an SMT solver.
\item Add auxiliary lemmas to assist in the proof.
\end{enumerate}

\subsection{Theory embedding}
Translating the generated verification conditions into the SMT-LIB input language requires first embedding the corresponding theory. The conditions are based on manipulating a program state object and are constructed as a sequence of function applications that modify the program state, together with formulas that refer to variable values extracted from it.

The program state type is defined as follows:
\begin{lstlisting}
(declare-sort Var)
(declare-sort Process)
(declare-sort PState)
(declare-datatypes ()
 ((State
    emptyState
    (toEnv (prev State))
    (setVarBool (prev State) 
        (varBool Var) (valBool Bool))
    ...
    (setPstate (prev State) 
        (proc Process)(pval PState))
    (reset (prev State) 
        (procReset Process))
 )))
\end{lstlisting}

For the getter functions, an uninterpreted function approach is used:
\begin{lstlisting}
(declare-fun getVarBool 
    (State String) Bool)
(assert
 (forall ((s State)(x Var)(v Bool))
   (= (getVarBool 
        (setVarBool s x v) x) 
    v)))
(assert
 (forall ((s State)(x Var)(Bool)(v Int))
   (=> (not (= x y))
       (= (getVarInt (setVarInt s x v) y)
          (getVarInt s y)))))
(assert
 (forall ((s State)(x Var)(b Int)(y Var))
   (= (getVarBool (setVarInt s x b) y)
      (getVarBool s y))))
\end{lstlisting}
Other getters are defined in a similar manner.

The \texttt{substate} function checks whether one state is a predecessor of another state:

\begin{lstlisting}
(define-fun-rec substate 
    ((s State) (t State)) Bool
 (match t
  ((emptyState
       (= s emptyState))
   ((toEnv s1)
       (or (= s t)
           (substate s s1)))
   ...
 ))
\end{lstlisting}

The function \texttt{toEnvP} checks whether a state is external:
\begin{lstlisting}
(define-fun toEnvP ((s State)) Bool
   (is-toEnv s))
\end{lstlisting}

The function \texttt{predEnv} returns the previous external state:
\begin{lstlisting}
(define-fun-rec predEnv ((s State)) State
 (match s
  ((emptyState emptyState))
  ((toEnv s1)
       (ite (is-toEnv s1)
            s1
            (predEnv s1)))
   ...
 ))
\end{lstlisting}

\subsection{Auxiliary lemmas}

To improve the likelihood of successful verification and to broaden the set of correctness conditions that can be proved, additional lemmas must be added. However, excessive expansion of the lemma pool can lead to a significant drop in performance. To avoid this, all auxiliary lemmas are divided into several categories:
\begin{enumerate}
\item \textbf{Base.} This group consists of lemmas describing simple properties of the program state datatype and its functions, and is universal to all verification conditions. For example, reflexivity of \texttt{substate}: \lstinline{(assert (forall ((s State)) (substate s s)))}
\item \textbf{Advanced.} This group consists of user-defined assumptions and invariants, as well as three kinds of extra invariants: process states, defined process state variables, and defined stabilized process state variables.
\item \textbf{Optional.} This group consists of the extra invariants for process state transition conditions.
\end{enumerate}

By default, only base and advanced lemmas are used. Optional lemmas must be enabled separately. Additionally, each verification condition interacts with at most two process states (the one in which it starts and the one in which it ends). These factors allow us to further narrow down the proof context by discarding advanced lemmas unrelated to the states involved.

\section{VCG modification}
\begin{figure}[h]
    \centering
    \includegraphics[scale=0.1]{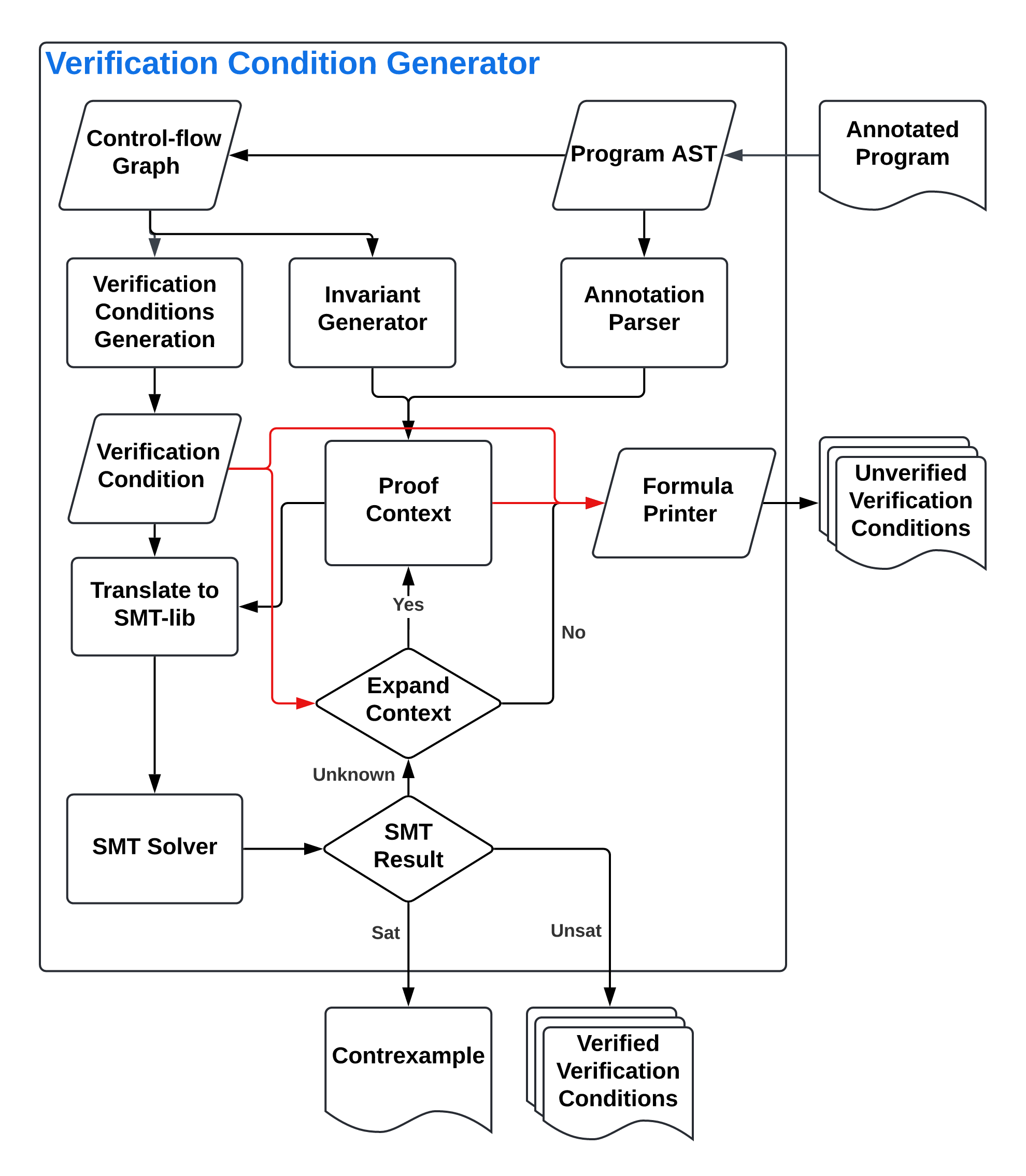}
    \caption{Modified algorithm of VCG.}
    \label{fig:algorithm}
\end{figure}

In \cite{ishchenko2025verification}, an algorithm for generating verification conditions was described. This algorithm builds a control-flow graph of the program and traverses it, constructing verification conditions as lists of formulas, which are then output as Isabelle/HOL formulas. The architecture of the modified verification condition generator (VCG) is shown in Fig.~\ref{fig:algorithm}.

The verification process begins by parsing the annotated program into an abstract syntax tree (AST). From this AST, a control-flow graph (CFG) of the program is constructed. In parallel with CFG construction, the annotation parser processes user-provided annotations, translates them into an internal logical representation, and associates them with the corresponding program statements and declarations.

The invariant generator analyzes the structure of the control-flow graph and produces additional invariants as described in Sect.~\ref{extra}. These invariants, together with the translated annotations, are used to form a proof context.

Next, the verification condition generation algorithm traverses the control-flow graph and produces VCs. Each condition is processed independently. For every generated VC, a corresponding proof context is constructed. This proof context contains auxiliary lemmas derived from annotations and generated invariants, as well as a lemma representing the behavioral requirement of the program. The negation of this requirement is used as the proof goal in the SMT query.

Before being submitted to the solver, the verification condition together with its proof context is translated into the SMT-LIB format. The result is then passed to an SMT solver (CVC5 by default).

If the solver returns \textit{sat}, the requirement is violated and a counterexample is produced.

If the solver returns \textit{unsat}, the verification condition is proven valid and added to the list of verified conditions.

If the solver returns \textit{unknown}, the system attempts to refine the proof context. In this case, additional lemmas may be introduced into the context, and the verification condition is translated again and resubmitted to the SMT solver. If further expansion is not possible or has been disabled, the original verification condition and its proof context are printed and added to the list of unverified verification conditions for manual proof in a theorem prover such as Isabelle/HOL.

\section{Conclusion}
\label{sec5}
This paper has presented an approach to automating the proof of verification conditions in the context of deductive verification for Reflex programs. The work addresses a critical bottleneck in the verification process: the large number of verification conditions generated for realistic industrial programs, which makes manual proof impractical.

We have proposed three complementary extensions to our previously developed verification condition generator. First, we introduced a redesigned annotation language that allows engineers to specify program properties in a structured and expressive manner, supporting both a default logical notation and integration with external languages such as Isabelle/HOL. Second, we developed a mechanism for automatically generating structure-based invariants derived from the control-flow graph of the program, including invariants describing process states, variable values in stabilized and non-stabilized states, and process state transition conditions. Third, we integrated SMT solvers into the verification workflow by translating verification conditions into the SMT-LIB format and incorporating a layered system of auxiliary lemmas to support automated reasoning.

The modified VCG architecture combines these elements into a unified verification pipeline. Starting from an annotated program, it constructs a control-flow graph, processes user annotations, generates additional invariants, and produces verification conditions. Each condition is then translated into an SMT query together with a tailored proof context. The solver's response determines whether the condition is proved, a counterexample is found, or further refinement is needed, with unverified conditions deferred to interactive theorem provers like Isabelle/HOL.

This work represents a step toward making deductive verification more practical for industrial process-oriented programs by reducing the manual effort required to prove verification conditions. The proposed approach leverages both user-provided specifications and automatically derived structural properties, combining them with the power of modern SMT solvers.

Future work will focus on the practical implementation and experimental evaluation of the proposed system. This includes implementing the annotation parser, invariant generator, and SMT-LIB translator, followed by a comprehensive evaluation on a suite of Reflex programs of varying complexity. We also plan to investigate additional sources of automatically generated invariants, refine the lemma selection strategy to improve solver performance, and explore techniques for handling more complex verification conditions that currently require interactive proof.

\bibliographystyle{IEEEtran}
\bibliography{lit.bib}

\end{document}